\documentclass[aps,prb,twocolumn,superscriptaddress,linenumbers]{revtex4}
\usepackage{wrapfig}
\usepackage{amsmath}
\usepackage{amssymb}
\usepackage{hyperref}
\usepackage{graphicx}
\usepackage{color}

\begin{document}


\title{Beyond the Jaynes-Cummings model: circuit QED in the ultrastrong coupling regime}




\author{T.~Niemczyk}
\affiliation{Walther-Mei{\ss}ner-Institut, Bayerische Akademie der
Wissenschaften, D-85748~Garching, Germany}\email[]{email: thomas.niemczyk@wmi.badw.de}

\author{F.~Deppe}
\affiliation{Walther-Mei{\ss}ner-Institut, Bayerische Akademie der
Wissenschaften, D-85748~Garching, Germany}
\affiliation{Physik-Department, Technische Universit\"{a}t
M\"{u}nchen, D-85748 Garching, Germany}
\author{H.~Huebl}
\affiliation{Walther-Mei{\ss}ner-Institut, Bayerische Akademie der
Wissenschaften, D-85748~Garching, Germany}
\author{E.~P.~Menzel}
\affiliation{Walther-Mei{\ss}ner-Institut, Bayerische Akademie der
Wissenschaften, D-85748~Garching, Germany}
\author{F.~Hocke}
\affiliation{Walther-Mei{\ss}ner-Institut, Bayerische Akademie der
Wissenschaften, D-85748~Garching, Germany}
\author{M.~J.~Schwarz}
\affiliation{Walther-Mei{\ss}ner-Institut, Bayerische Akademie der
Wissenschaften, D-85748~Garching, Germany}
\author{J.~J.~Garcia-Ripoll}
\affiliation{Instituto de F\'{i}sica Fundamental, CSIC, Serrano
113-bis, 28006 Madrid, Spain}
\author{D.~Zueco}
\affiliation{Instituto de Ciencia de Materiales de Arag\'on y
Departamento de F\'{\i}sica de la Materia Condensada,
CSIC-Universidad de Zaragoza, E-50012 Zaragoza, Spain.}
\author{T.~H\"{u}mmer}
\affiliation{Institut f\"ur Physik, Universit\"at Augsburg,
Universit\"atsstra{\ss}e~1, D-86135 Augsburg, Germany}
\author{E.~Solano}
\affiliation{Departamento de Qu\'{\i}mica F\'{\i}sica, Universidad
del Pa\'{\i}s Vasco - Euskal Herriko Unibertsitatea, Apdo. 644,
48080 Bilbao, Spain} \affiliation{IKERBASQUE, Basque Foundation
for Science, Alameda Urquijo 36, 48011 Bilbao, Spain}
\author{A.~Marx}
\affiliation{Walther-Mei{\ss}ner-Institut, Bayerische Akademie der
Wissenschaften, D-85748~Garching, Germany}
\author{R.~Gross}
\affiliation{Walther-Mei{\ss}ner-Institut, Bayerische Akademie der
Wissenschaften, D-85748~Garching, Germany}
\affiliation{Physik-Department, Technische Universit\"{a}t
M\"{u}nchen, D-85748 Garching, Germany}

\date{\today}

\begin{abstract}
In cavity quantum electrodynamics (QED)~\cite{Mabuchi:2002a,Walther:2006a,Haroche:2006a}, light-matter interaction is probed at its most fundamental level, where individual atoms are coupled to single photons stored in three-dimensional cavities. This unique possibility to experimentally explore the foundations of quantum physics has greatly evolved with the advent of circuit QED~\cite{Blais:2004a,Wallraff:2004a,Chiorescu:2004a,Johansson:2006a,Schuster:2007a,Astafiev:2007a,Deppe:2008a,Fink:2008a,Abdumalikov:2008,Hofheinz:2009a}, where on-chip superconducting qubits and oscillators play the roles of two-level atoms and cavities, respectively. In the strong coupling limit, atom and cavity can exchange a photon frequently before coherence is lost. This important regime has been reached both in cavity and circuit QED, but the design flexibility and engineering potential of the latter allowed for increasing the ratio between the atom-cavity coupling rate $g$ and the cavity transition frequency $\omega_{\rm r}$ above the percent level~\cite{Schuster:2007a,Schoelkopf:2008a,Bishop:2008a}. While these experiments are well described by the renowned Jaynes-Cummings model~\cite{JaynesCummings:1963a}, novel physics is expected when $g$ reaches a considerable fraction of $\omega_{\rm r}$. Promising steps towards this so-called ultrastrong coupling regime~\cite{Ciuti:2006a,Devoret:2007a} have recently been taken in semiconductor structures~\cite{Guenter:2009a,Anappara:2009a}. Here, we report on the first experimental realization of a superconducting circuit QED system in the ultrastrong coupling limit and present direct evidence for the breakdown of the Jaynes-Cummings model. We reach remarkable normalized coupling rates $g/\omega_{\rm r}$ of up to 12\,\% by enhancing the inductive coupling of a flux qubit~\cite{Mooij:1999a} to a transmission line resonator using the nonlinear inductance of a Josephson junction~\cite{Bourassa:2009a}. Our circuit extends the toolbox of quantum optics on a chip towards exciting explorations of the ultrastrong interaction between light and matter.
\end{abstract}
\maketitle In the strong coupling regime, the atom-cavity coupling rate $g$ exceeds the dissipation rates $\kappa$ and $\gamma$ of both, cavity and atom, giving rise to coherent light-matter oscillations and superposition states. This regime was reached in various types of systems operating at different energy scales~\cite{Rempe:1992a,Mabuchi:2002a,Walther:2006a,Haroche:2006a,Reithmaier:2004a,Groebelbacher:2009a}. At microwave frequencies, strong coupling is feasible due to the enormous engineerability of superconducting circuit QED systems~\cite{Blais:2004a,Wallraff:2004a}. Here, small cavity mode volumes and large dipole moments of artificial atoms~\cite{Niemczyk:2009a} enable coupling rates $g$ of about~\cite{Bishop:2008a} 1\,\% of the cavity mode frequency $\omega_{\rm r}$. Nevertheless, as in cavity QED, the quantum dynamics of these systems follows the Jaynes-Cummings model, which describes the coherent exchange of a single excitation between the atom and the cavity mode. Although the Hamiltonian of a realistic atom-cavity system contains so-called counterrotating terms allowing the simultaneous creation ior annihilation of an excitation in both atom and cavity mode, these terms can be safely neglected for small normalized coupling rates $g/\omega_{\rm r}$. However, when $g$ becomes a significant fraction of $\omega_{\rm r}$, the counterrotating terms are expected to manifest, giving rise to exciting effects in QED.

The ultrastrong coupling regime is difficult to reach in traditional quantum optics, but was recently realized in a solid-state semiconductor system~\cite{Guenter:2009a,Anappara:2009a}. There, quantitative deviations from the Jaynes-Cummings model have been observed, but a direct experimental proof of its breakdown by means of an unambiguous feature is still missing. In this report, we exploit the potential of flux-biased superconducting quantum circuits to reach the ultrastrong coupling regime~\cite{Devoret:2007a,Bourassa:2009a}. For this purpose, we increase $g/\omega_{\rm r}$ up to 12\,\% utilizing the large nonlinear inductance of a Josephson junction (JJ) shared between a flux qubit and a coplanar waveguide resonator. We explicitly make use of the multimode structure of our resonator, allowing the direct observation of physics beyond the Jaynes-Cummings model. In equilibrium, our transmission spectra reveal anticrossings which can be clearly attributed to the counterrotating terms in the system Hamiltonian. These anticrossings are caused by the simultaneous creation (annihilation) of two excitations, one in the qubit and one in a resonator mode, while annihilating (creating) only one excitation in a different resonator mode.
\begin{figure*}[tb]
\includegraphics[width=0.7\linewidth]{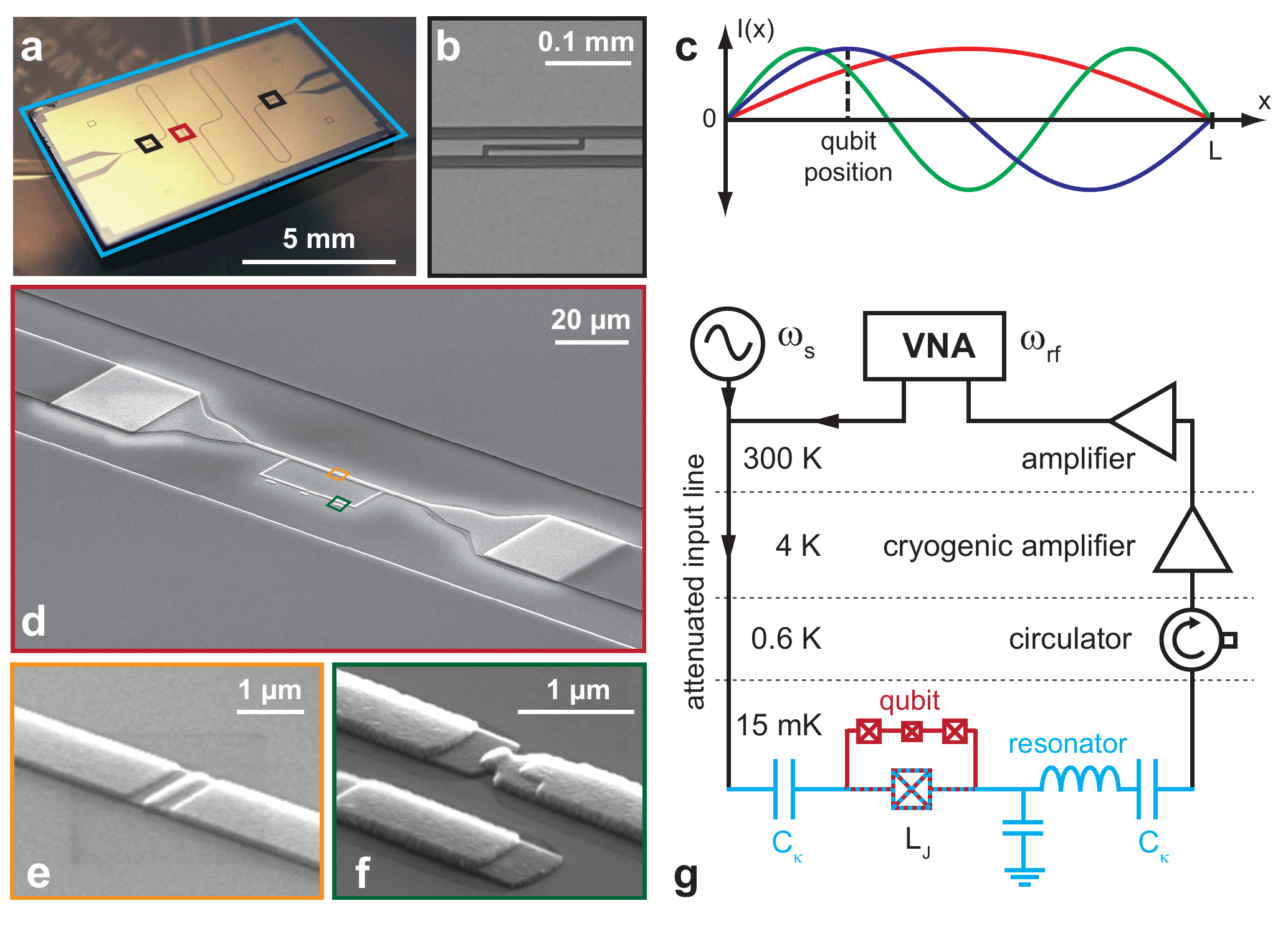}
\caption{\textbf{Quantum circuit and experimental setup.}
\textbf{a}, Optical image of the superconducting $\lambda/2$
coplanar waveguide resonator (light blue box). Black rectangles: area shown in \textbf{b}. Red rectangle: area shown in \textbf{d}. \textbf{b}, SEM image of one of the coupling capacitors. \textbf{c}, Sketch of the current distribution of the first three resonator modes. Their resonance frequencies are $\omega_{\rm 1}/2\pi = 2.782$\,GHz ($\lambda/2$, red), $\omega_{\rm 2}/2\pi = 5.357$\,GHz ($\lambda$, blue), and $\omega_{\rm 3}/2\pi = 7.777$\,GHz ($3\lambda/2$, green). \textbf{d}, SEM image of the galvanically coupled flux qubit. The fabrication technology for qubit and resonator is described elsewhere~\cite{Niemczyk:2009a}. The width of the center conductor is $20\,\mu$m, that of the constriction $1\,\mu$m. The area of the qubit loop is 180~$\mu\text{m}^2$. Orange rectangle: area shown in \textbf{e}. Green rectangle: area shown in \textbf{f}. \textbf{e}, SEM image of the large JJ with a Josephson inductance $L_{\rm J}$, whose large inductance is responsible for approximately 85\,\% of the qubit-resonator coupling. \textbf{f}, One JJ of the qubit loop. The area of this junction is 14\,\% of the one shown in \textbf{e}. \textbf{g}, Schematic sketch of the measurement setup. The transmission through the cavity at $\omega_{\rm rf}$ is measured using a VNA. A second microwave signal at $\omega_{\rm s}$ is used for two-tone qubit spectroscopy. The input signal is attenuated at various temperature stages and coupled into the resonator (light blue) via the capacitors $C_{\rm \kappa}$. The crossed boxes represent Josephson junctions. A circulator isolates the sample from the amplifier noise.}\label{FIG1}
\end{figure*}
\begin{figure*}[tb]
\includegraphics[width=0.9\linewidth]{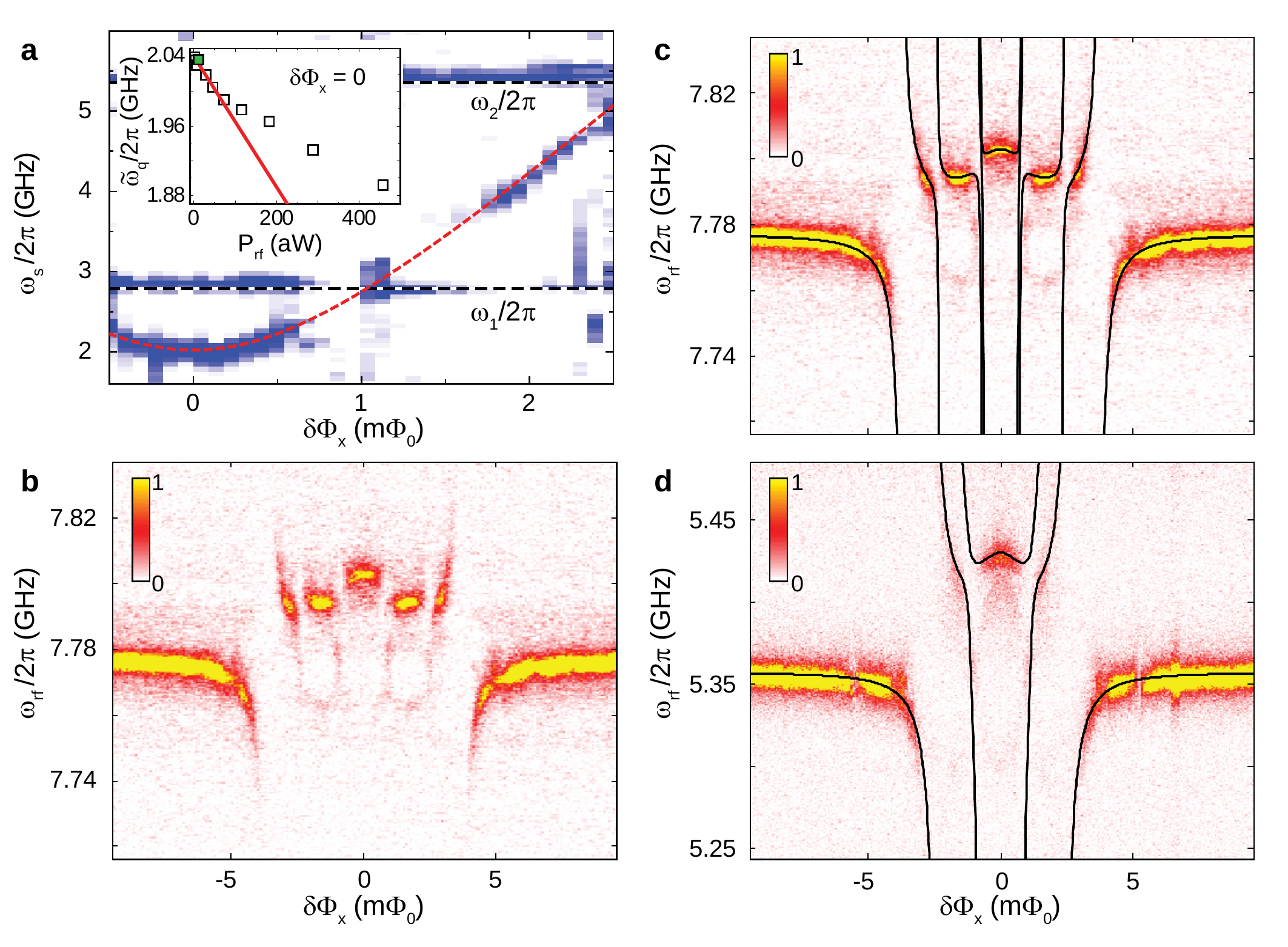}
\caption{\textbf{Qubit microwave spectroscopy and low power
transmission spectra.} \textbf{a}, Microwave spectroscopy of the coupled qubit-cavity system. The measured transmission magnitude (color coded, blue: low; white: high) is plotted as a function of the relative flux bias $\delta\Phi_{\rm x}$ and the spectroscopy frequency $\omega_{\rm s}/2\pi$. The red broken line indicates the dressed qubit transition frequency $\tilde{\omega}_{\rm q}$\cite{Schuster:2005a,Abdumalikov:2008}. Inset: center frequency of the qubit spectroscopy signal at $\delta\Phi_{\rm x} = 0$ as a function of the probe power $P_{\rm rf }$. The FWHM of the qubit signal is approximately 80 MHz in the low power limit $P_{\rm rf}\,P_{\rm s} \rightarrow 0$. Red line: fit to the linear region~\cite{Zueco:2009a}. The green dot indicates the power
level at which the spectra in \textbf{b}, \textbf{c} and
Fig.~\ref{FIG3} are recorded. \textbf{b}, Cavity transmission
($3\lambda/2$-mode, linear scale, arb. units) as a function of
$\delta\Phi_{\rm x}$ and probe frequency $\omega_{\rm rf}/2\pi$. \textbf{c}, Same spectrum as in \textbf{b}. Black lines: numerical fit of the spectrum of the Hamiltonian~(\ref{FullH}) to the data. \textbf{d}, Cavity transmission ($\lambda$-mode, linear scale, arb.~units) as a function of $\delta\Phi_{\rm x}$ and probe frequency $\omega_{\rm rf}/2\pi$. The spectrum is recorded at $P_{\rm rf}$ corresponding to $\bar{n}_{\rm 2} \approx 0.9$ due to a higher insertion loss of this cavity mode. Black lines: numerically evaluated energy level spectrum with parameters from \textbf{c}.}
\label{FIG2}
\end{figure*}
\begin{figure*}[htb]
\includegraphics[width=0.9\linewidth]{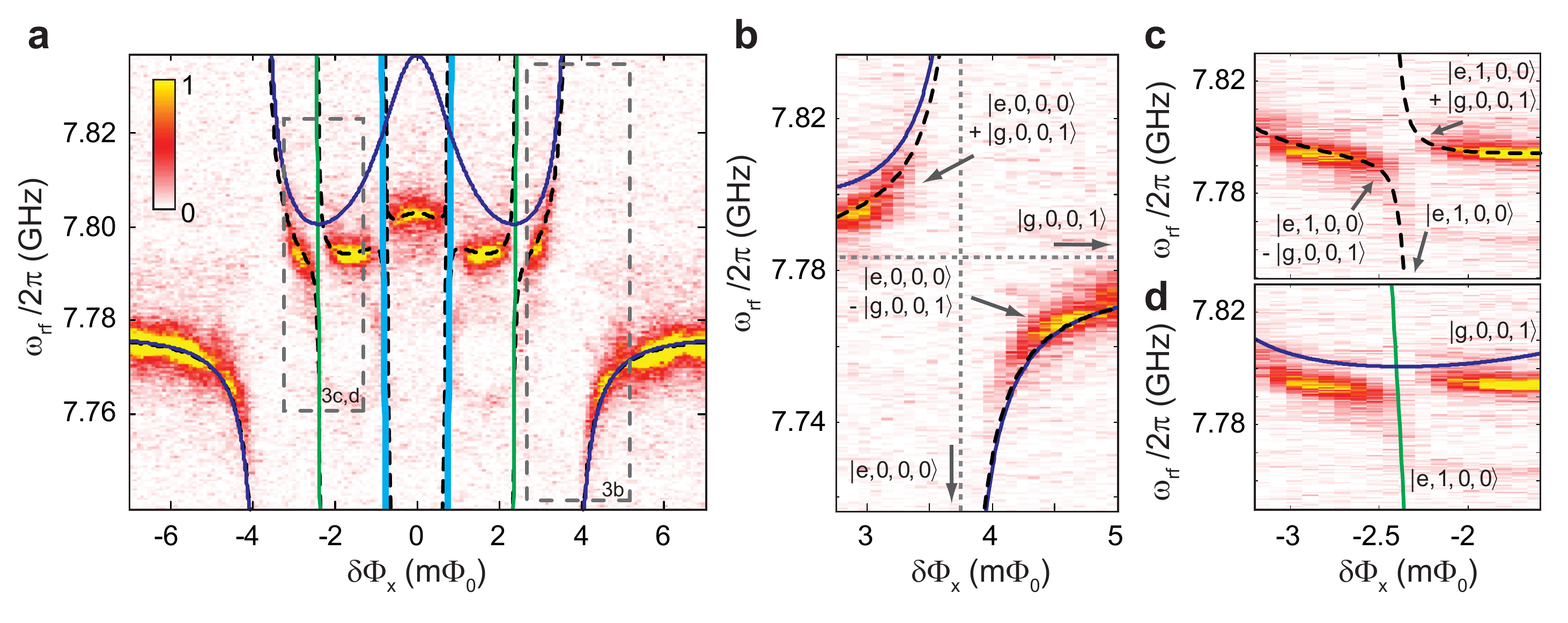}
\caption{\textbf{Breakdown of the Jaynes-Cummings model.}
\textbf{a}, Cavity transmission ($3\lambda/2$-mode, linear scale, arb.~units) as a function of $\delta\Phi_{\rm x}$ and probe frequency $\omega_{\rm rf}/2\pi$. Black broken lines in all plots: energy level spectrum obtained from the Hamiltonian~(\ref{FullH}). Colored lines in all plots: energy level spectrum obtained from the Jaynes-Cummings model (dark blue: $|g,0,0,1\rangle$, except for anticrossing region shown in \textbf{b}; green: $|e,1,0,0\rangle$; light blue: $|e,0,1,0\rangle$ and $|e,2,0,0\rangle$, which are indistinguishable within the resolution of this plot. Grey broken boxes: areas magnified in \textbf{b}-\textbf{d}. \textbf{b}, Single excitation anticrossing. The quantitative deviations of the Jaynes-Cummings model from Eq.~(\ref{FullH}) are attributed to a small admixture of the state $|g,1,1,0\rangle$. The uncoupled states are indicated
by grey broken lines. \textbf{c}, Avoided crossing due to a
coupling between the degenerate states $|g,0,0,1\rangle$ and
$|e,1,0,0\rangle$. This is caused by counterrotating terms in the Hamiltonian~(\ref{FullH}). A detailed analysis yields a minor admixture of $|g,1,1,0\rangle$ (superposition states: $\approx |g,0,0,1\rangle/\sqrt{2} \pm (|e,1,0,0\rangle/\sqrt{3} + |g,1,1,0\rangle/\sqrt{6}$)). This admixture has no effect on the reasoning presented in the main text. The energy level spectrum obtained from the Jaynes-Cummings model is omitted for clarity. \textbf{d}, Same as \textbf{c}, but for the Jaynes-Cummings model. Within numerical accuracy, no anticrossing is predicted, clearly contradicting the data.} \label{FIG3}
\end{figure*}

Images of our quantum circuit and a schematic of the measurement setup are shown in Figure \ref{FIG1}. At a current antinode for the $\lambda$-mode of a niobium superconducting resonator (Fig.~\ref{FIG1}a-c), a part of the center conductor is replaced with a narrow aluminum strip interrupted by a large-area JJ. This junction mediates most of the inductive coupling between a superconducting flux qubit~\cite{Mooij:1999a} galvanically connected to the strip. The qubit consists of three nanometer-scaled JJs interrupting a superconducting loop, which is threaded by an external flux bias $\Phi_{\rm x}$. Scanning electron microscope (SEM) images of the qubit loop and the JJs are shown in Figure~\ref{FIG1}d-f. For suitable junction sizes, the qubit potential landscape can be reduced to a double-well potential, where the two minima correspond to states with clockwise and counter-clockwise persistent currents $|\pm I_{\rm p}\rangle$. At $\delta\Phi_{\rm x} \,=\, \Phi_{\rm x} - \Phi_{\rm 0}/2 = 0$, these two states are degenerate and separated by an energy gap $\Delta$. In the qubit eigenbasis, the qubit Hamiltonian reads $\hat{H}_{\rm q} \,=\, \hbar\omega_{\rm q}\hat{\sigma}_{\rm z}/2$. Here, $\omega_{\rm q} \,=\, \sqrt{\Delta^2 + (2I_{\rm p}\cdot\delta\Phi_{\rm x})^2}/\hbar$ is the qubit transition frequency which can be adjusted by an external flux bias. We note, that for our flux qubit the two-level approximation is well justified due to its large anharmonicity. The resonator modes are described as harmonic oscillators, $\hat{H}_{\rm n} \,=\, \hbar\omega_{\rm n}(\hat{a}_{\rm n}^{\dagger}\hat{a}_{\rm n} \,+\, 1/2)$, where $\omega_{\rm n}$ is the resonance frequency and $n$ is the resonator mode index. The operator $\hat{a}_{\rm n}^{\dagger}$ ($\hat{a}_{\rm n}$) creates (annihilates) a photon in the $n$th resonator mode. Due to the inhomogeneous transmission line geometry~\cite{Bourassa:2009a} (see Fig. \ref{FIG1}d), the higher mode frequencies of our resonator are not integer multiples of the fundamental resonance frequency $\omega_{\rm 1}$ . Throughout this work, we refer to the $n$th mode as the $n\lambda/2$-mode. Then, the Hamiltonian of our our quantum circuit can be written as
\begin{equation}
\hat{H}  =  \hat{H}_{\rm q}  +  \sum\limits_{n} \left[\hat{H}_{\rm
n} + \hbar g_{\rm n}\left(\hat{a}_{\rm n}^{\dagger}+\hat{a}_{\rm
n} \right)\left(\cos\theta\,\hat{\sigma}_{\rm z} -
\sin\theta\,\hat{\sigma}_{\rm x}\right)\right].\label{FullH}
\end{equation}
Here, $\hat{\sigma}_{\rm x,z}$ denote Pauli operators, $g_{\rm n}$ is the coupling rate of the qubit to the $n$th cavity mode, and the flux dependence is encoded in $\sin\theta = \Delta/\hbar\omega_{\rm q}$ and $\cos\theta$. The operator $\hat{\sigma}_{\rm x}$ is conveniently expressed as sum of the qubit raising ($\hat{\sigma}_{\rm +}$) and lowering ($\hat{\sigma}_{\rm -}$) operator. Thus, in contrast to the Jaynes-Cummings model, the Hamiltonian in Eq.~(\ref{FullH}) explicitly contains counterrotating terms of the form $\hat{a}_{\rm n}^{\dagger}\hat{\sigma}_{\rm +}$ and $\hat{a}_{\rm n}\hat{\sigma}_{\rm -}$.

Figure \ref{FIG1}g shows a schematic of our measurement setup. The quantum circuit is located at the base temperature of 15~mK in a dilution refrigerator. We measure the amplified resonator transmission using a vector network analyzer (VNA). For qubit spectroscopy measurements, the system is excited with a second microwave tone $\omega_{\rm s}$ with power $P_{\rm s}$, while using the $3\lambda/2$-mode at $\omega_{\rm 3}/2\pi = 7.777$\,GHz for dispersive readout~\cite{Schuster:2005a,Abdumalikov:2008}.

We first present measurements allowing the extraction of the coupling constants of the qubit to the first three resonator modes. The spectroscopy data in Fig.~\ref{FIG2}a shows the dressed qubit transition frequency~\cite{Blais:2004a,Schuster:2005a} with the expected hyperbolic flux-dependence and a minimum at $\delta\Phi_{\rm x} = 0$. Furthermore, flux-independent features corresponding to the two lowest resonator modes ($\omega_{\rm 1}$ and $\omega_{\rm 2}$) are visible. In principle, a fit to the Hamiltonian in Eq.~(\ref{FullH}) would yield all system parameters. However, our measurement resolution does not allow us to reliably determine the coupling constants $g_{\rm n}$ in this situation. Instead, we extract $g_{\rm n}$ from a cavity transmission spectrum with negligible photon population. For that purpose, we first measure the power-dependent ac-Zeeman shift of the qubit transition frequency at $\delta\Phi_{\rm x}\,=\,0$. The data is shown in the inset of Fig.~\ref{FIG2}a. The average photon number $\bar{n}_{\rm 3}$ can be estimated using the relation $P_{\rm rf} = \bar{n}_{\rm 3}\hbar\omega_{\rm 3}\kappa_{\rm 3}$~\cite{Astafiev:2007a,Fink:2008a}, where $\kappa_{\rm 3}/2\pi \approx 3.7$ MHz is the full width at half maximum (FWHM) of the cavity resonance and $P_{\rm rf}$ the probe power referred to the input of the resonator. Figure~\ref{FIG2}b shows a color coded transmission spectrum for the $3\lambda/2$-mode as a function of $\delta\Phi_{\rm x}$. The data is recorded at an input power $P_{\rm rf} \approx -140$~dBm (green data point in Fig.~\ref{FIG2}a, inset) corresponding to $\bar{n}_{\rm 3} = 0.18$.

We observe a spectrum with a large number of anticrossings resulting from the multimode structure of our cavity system. To extract the individual coupling constants $g_{\rm n}$, we compute the lowest nine transition frequencies of the Hamiltonian given in Eq.~(\ref{FullH}) incorporating the first three resonator modes. Fitting the results to the spectrum of the $3\lambda/2$-mode shows excellent agreement with the measured data as shown in Fig.~\ref{FIG2}c. We note that the spectrum for the $\lambda$-mode shown in Fig.~\ref{FIG2}d can be well described without additional fitting using the parameters extracted from the $3\lambda/2$-mode. For the qubit, we obtain $\Delta/h = 2.25$~GHz and $2I_{\rm p} = 630$~nA. Most importantly, we find coupling rates of $g_{\rm 1}/2\pi = 314$~MHz, $g_{\rm 2}/2\pi = 636$~MHz, and $g_{\rm 3}/2\pi = 568$~MHz. The values for $g_{\rm n}$ correspond to normalized coupling rates $g_{\rm n}/\omega_{\rm n}$ of remarkable 11.2\,\%, 11.8\,\%, and 7.3\,\%, respectively. From these numbers, we expect significant deviations of our system from a three-mode Jaynes-Cummings model, ultimately proving that we have successfully reached the ultrastrong coupling regime.

In the following, we analyze the features in our data which constitute unambiguous evidence for the breakdown of the rotating-wave approximation inherent to the Jaynes-Cummings model. In Figure~\ref{FIG3}, we compare the energy level spectrum of the Hamiltonian in Eq.~(\ref{FullH}) to that of a three-mode Jaynes-Cummings model. We note that, depending on $\delta\Phi_{\rm x}$, there are regions where our data can be well described by the Jaynes-Cummings model, and regions where there are significant deviations (see Fig.~\ref{FIG3}a). For our analysis we use the notation $|q,N_{\rm 1},N_{\rm 2},N_{\rm 3}\rangle = |q\rangle \otimes |N_{\rm 1}\rangle \otimes |N_{\rm 2}\rangle \otimes |N_{\rm 3}\rangle$, where $q = \{g,e\}$ denote the qubit ground or excited state, respectively, and $|N_{\rm n}\rangle = \{|0\rangle,|1\rangle,|2\rangle,\dots\}$ represents the Fock-state with photon occupation $N$ in the $n$th resonator mode. At the outermost anticrossings (Fig.~\ref{FIG3}b), where $\omega_{\rm 3} \approx \omega_{\rm q}$, the eigenstates $|\psi_{\pm }\rangle$ of the coupled system are in good approximation symmetric and antisymmetric superpositions of $|e,0,0,0\rangle$ and $|g,0,0,1\rangle$. This exchange of a single excitation between qubit and resonator is a characteristic of the Jaynes-Cummings model. On the contrary, the origin of the anticrossing shown in Fig.~\ref{FIG3}c is of different nature: the dominant contributions to the eigenstates $|\psi_{\pm}\rangle$ are approximate symmetric and antisymmetric superpositions of the degenerate states $\varphi_{\rm 1} = |e,1,0,0\rangle$ and $\varphi_{\rm 2} = |g,0,0,1\rangle$. The transition from $\varphi_{\rm 1}$ to $\varphi_{\rm 2}$ can be understood as the annihilation of two excitations, one in the $\lambda/2$-mode \emph{and} one in the qubit, while, simultaneously, creating only one excitation in the $3\lambda/2$-mode. Such a process can only result from counterrotating terms as they are present in the Hamiltonian (\ref{FullH}), but not within the Jaynes-Cummings approximation. Here, only eigenstates with an equal number of excitations can be coupled. Although counterrotating terms in principle exist in any real circuit QED system, their effects become prominent only in the ultrastrong coupling limit with large normalized couplings $g_{\rm n}/\omega_{\rm n}$ as realized in our system. Hence, the observed anticrossing shown in Fig.~\ref{FIG3}c is a direct experimental manifestation of physics beyond the rotating-wave approximation in the Jaynes-Cummings model. As shown in Fig.~\ref{FIG3}d, the latter would imply a crossing of the involved energy levels, which is not observed. A similar argument applies to the innermost anticrossings (see Fig.~\ref{FIG3}a), although the involved eigenstates have a more complicated character.

In conclusion, we present measurements on a flux-based superconducting circuit QED system in the ultrastrong coupling regime. The results are in excellent agreement with theoretical predictions and show clear evidence for physics beyond the Jaynes-Cummings model. Our system can act as an on-chip prototype for unveiling the physics of ultrastrong light-matter interaction. Future explorations may include squeezing, switchable ultrastrong coupling~\cite{Peropadre:2009a}, causality effects in quantum field theory~\cite{Sabin:2009a}, and the generation of bound states of qubits and photons.

\section*{Acknowledgments}

We thank G. M. Reuther for discussions and T. Brenninger, C.
Probst, and K. Uhlig for technical support. We acknowledge financial support by the Deutsche Forschungsgemeinschaft via SFB~631 and the German Excellence Initiative via NIM. E.S. acknowledges funding from UPV/EHU Grant GIU07/40, Ministerio de Ciencia e Innovaci\'on FIS2009-12773-C02-01, European Projects EuroSQIP and SOLID. D.Z. acknowledges financial support from FIS2008-01240 and FIS2009-13364-C02-0 (MICINN).

\section*{Author contributions}
T.N. fabricated the sample, conducted the experiment and analyzed the data presented in this work. F.D. provided important contributions regarding the interpretation of the results. T.N. and F.D. co-wrote the manuscript. J.J.G.-R. provided the basic idea and the techniques for the numerical analysis of the data. E.S. and J.J.G.-R. supervised the interpretation of the data. D.Z. and T.H. contributed to the understanding of the results and developed an
analytical model of our system. H.H. contributed to the numerical analysis and helped with the experiment. E.P.M. contributed strongly to the experimental setup. M.J.S. and F.H. contributed to discussions and helped editing the manuscript. A.M. and R.G. supervised the experimental part of the work.

\bibliography{myBib}

\end{document}